\documentclass[12pt]{article}\usepackage[]{graphicx}\usepackage[]{color}
\makeatletter
\def\maxwidth{ %
  \ifdim\Gin@nat@width>\linewidth
    \linewidth
  \else
    \Gin@nat@width
  \fi
}
\makeatother

\definecolor{fgcolor}{rgb}{0.345, 0.345, 0.345}
\makeatletter
\@ifundefined{AddToHook}{}{\AddToHook{package/xcolor/after}{\definecolor{fgcolor}{rgb}{0.345, 0.345, 0.345}}}
\makeatother

\usepackage{framed}
\makeatletter
 {\par\unskip\endMakeFramed%
 \at@end@of@kframe}
\makeatother

\definecolor{shadecolor}{rgb}{.97, .97, .97}
\definecolor{messagecolor}{rgb}{0, 0, 0}
\definecolor{warningcolor}{rgb}{1, 0, 1}
\definecolor{errorcolor}{rgb}{1, 0, 0}
\makeatletter
\@ifundefined{AddToHook}{}{\AddToHook{package/xcolor/after}{
\definecolor{shadecolor}{rgb}{.97, .97, .97}
\definecolor{messagecolor}{rgb}{0, 0, 0}
\definecolor{warningcolor}{rgb}{1, 0, 1}
\definecolor{errorcolor}{rgb}{1, 0, 0}
}}
\makeatother
\newenvironment{knitrout}{}{} 

\usepackage{alltt}

\newcommand{\blind}{1}

\usepackage{xr-hyper}       
\usepackage{hyperref}
\usepackage{amsmath}

\usepackage{graphicx}
\usepackage{enumerate}
\usepackage{natbib}

\usepackage{etoolbox} 
\usepackage{amsthm, amssymb}
\usepackage{booktabs}       
\usepackage{algorithm}      
\usepackage{algpseudocode}
\usepackage{multirow}
\usepackage{sectsty}
\usepackage{tabularx}
\usepackage{graphicx}
\usepackage{tikz}           
\usepackage{bm}             
\usepackage{xcolor}
\usepackage{microtype} 
\usepackage{import}
\usepackage{url}
\usepackage{soul, color}
\usepackage{titling}
\usepackage{dsfont} 
\usetikzlibrary{arrows, snakes, backgrounds, patterns, matrix, shapes, fit, 
calc, shadows, plotmarks}

\graphicspath{{./figures/}}

\usepackage{multibib}
\newcites{App}{References}

\newcommand{\eat}[1]{}

\def\spacingset#1{\renewcommand{\baselinestretch}%
  {#1}\small\normalsize} \spacingset{1}

\newtheorem{definition}{Definition}

\usepackage{lscape}

\IfFileExists{upquote.sty}{\usepackage{upquote}}{}
\begin{document}

\def\spacingset#1{\renewcommand{\baselinestretch}%
{#1}\small\normalsize} \spacingset{1}

\AtBeginEnvironment{tabular}{\footnotesize}

\makeatletter
\AtBeginEnvironment{tabular}{%
  \def\baselinestretch{1}\@currsize}%
\makeatother


\if1\blind
{
  \title{\bf A Practical Approach to Proper Inference with Linked Data}
  \author{Andee Kaplan\textsuperscript{a} \and
	Brenda Betancourt\textsuperscript{b} \and 
	Rebecca C.~Steorts\textsuperscript{c}
}
\date{
	\textsuperscript{a}Department of Statistics, Colorado State University\\
	\textsuperscript{b}Department of Statistics, University of Florida\\
	\textsuperscript{c}Departments of Statistical Science and Computer Science, Duke University\\[2ex]
	\today}
  
  \maketitle
} \fi

\if0\blind
{
  \bigskip
  \bigskip
  \bigskip
  \begin{center}
    {\LARGE\bf A Practical Approach to Proper Inference with Linked Data}
\end{center}
  \medskip
} \fi

\bigskip

\begin{abstract}
Entity resolution (ER), comprising record linkage and de-duplication, is the process of merging noisy databases in the absence of unique identifiers to remove duplicate entities. One major challenge of analysis with linked data is identifying a representative record among determined matches to pass to an inferential or predictive task, referred to as the \emph{downstream task}. Additionally, incorporating uncertainty from ER in the downstream task is critical to ensure proper inference. To bridge the gap between ER and the downstream task in an analysis pipeline, we propose five methods to choose a representative (or \emph{canonical}) record from linked data, referred to as \emph{canonicalization}. Our methods are scalable in the number of records, appropriate in general data scenarios, and provide natural error propagation via a Bayesian canonicalization stage. The proposed methodology is evaluated on three simulated data sets and one application -- determining the relationship between demographic information and party affiliation in voter registration data from the North Carolina State Board of Elections. We first perform Bayesian ER and evaluate our proposed methods for canonicalization before considering the downstream tasks of linear and logistic regression. Bayesian canonicalization methods are empirically shown to improve downstream inference in both settings through prediction and coverage. 

\end{abstract}

\noindent%
\vfill

\newpage
\spacingset{1.45} 

\section{Introduction}
\label{sec:intro}

%

In many practical problems analysts seek to remove duplicate records across multiple noisy databases, a process known as entity resolution (ER). This is just one important task in ``data cleaning'' or ``data integration," where the outputs are then used for inferential and predictive analyses in areas of application such as government statistics, human rights, economics, precision medicine, and others \citep{christen_data_2012, Ilyas2019, christophides2020, binette2021entity, papadakis2021four}. This is just one stage of an \emph{analysis pipeline}, typically consisting of four data cleaning stages prior to the inferential task \citep{christen_data_2012, hogan2013quality, Herzog_2007, AbelDasylva2016, christen2019data, OHare2019, Vidhya2019, papadakis2021four},
\begin{align}
\label{eqn:data-clean}
\text{schema alignment} \rightarrow \text{blocking} \rightarrow \text{ER} \rightarrow \text{canonicalization} \rightarrow \text{analysis}
\end{align}
Equation~\ref{eqn:data-clean} shows a potential pipeline, including the cleaning stages of (1) schema alignment, (2) blocking, (3) ER, and (4) canonicalization. The first stage, known as ``schema alignment," entails identifying available information across multiple data sources for the purpose of joining these together. In the second stage, ``blocking," the goal is to place similar records together as a computational tool to speed finding matches. In the third stage, known as ER, one identifies duplicate records and outputs clusters of records that belong to the same partition. Similar records that are considered to represent the same latent entity are grouped together. The ER result can be represented as links between records that belong to the same cluster (the \emph{linkage structure}) or, equivalently, a partition of records into clusters. The resulting data set that is created as a result of ER is generally referred to as \emph{linked data}. The fourth stage, canonicalization, is the subject of this work. Canonicalization is an optional stage, depending on the inferential goal of an analysis. The partition of records that is the output of ER is used to create a representative record for each cluster. This process creates a representative data set, which is used as a set of inputs for the inferential analysis.


While removing duplicate entities through ER and canonicalization is worthy as a stand-alone data-cleaning goal, most analyses are primarily motivated by the performance of inferential or predictive analyses on the linked data (e.g., regression or classification). Going forward, we will denote any such post-linkage analysis as the \emph{downstream task}. In an ideal scenario, all records would be correctly clustered by the linkage structure that results from an ER model. In practice, however, ER error is a common occurrence and methods to propagate this error throughout subsequent analysis stages are essential. The primary contribution of this work is the development of a scalable, multi-stage approach, where the ER error is propagated through the downstream task via a stage that we refer to as \emph{posterior canonicalization}.

Our multi-stage approach handles the ER and \emph{canonicalization} stages separately from the downstream task and involves using the resulting clusters from the ER model to construct a data set (without duplicate information) comprised of the most representative or \emph{canonical} set of records. Undoubtedly, the quality of the linkage directly affects the canonicalization step and subsequent downstream tasks. Therefore, the construction of this canonical data set plays a key role in performing inferential downstream tasks -- such as regression analysis -- with the linked data.

\subsection{Motivating Example}
\label{voter-data}

Our work is motivated by voter registration data published by the North Carolina State Board of Elections (NCSBE). The NCSBE provides regular updates of their database, releasing snapshots of their registration database for archival purposes \citep{ncsbe}. Each snapshot contains identifying information about voters such as first name, middle name, last name, full address, city, state, date of birth, state of birth, and phone number. Unlike curated versions of this data \citep{christen_preparation_2014}, the NCSBE snapshots contain voter party registration status, which can be useful for inferential or predictive tasks. In addition, given the updating snapshot structure of the NCSBE, individuals are duplicated between (and potentially within) each data set. For example, an individual can be duplicated when they move or when they perform a name change. We consider five snapshots taken between 2018 and 2019, and refer to this collection as the North Carolina Voter Data (NCVD).

Table \ref{tab:ex-records} illustrates an individual who is believed to be duplicated in the NCVD five times. One goal of the analysis is to predict political affiliation using demographic information, such as race, sex and age. Unfortunately, it's not clear which record in Table \ref{tab:ex-records} most accurately reflects voter ``Mark Baker". We will attempt to overcome this issue using canonicalization. The process of canonicalization involves finding the ``best" representation of this voter's data fields in order to predict party affiliation. Multiple definitions of ``best" lead to multiple methods for performing canonicalization and we detail some of these choices in Section \ref{sec:canonicalization}.

\begin{table}

\caption{\label{tab:ex-records}\label{tab:ex-records} Five records that represent the same voter in the NCSBE data set.}
\centering
\begin{tabular}[t]{llllrl}
\toprule
Last & First & Race & Sex & Age & Party\\
\midrule
BAKER & MARK & WHITE & MALE & 42 & REP\\
BAKER & MARK & WHITE & MALE & 41 & DEM\\
BAKER & MARK & WHITE & MALE & 41 & REP\\
BAKER & MARK & WHITE & MALE & 40 & DEM\\
BAKER & MARK & WHITE & MALE & 40 & REP\\
\bottomrule
\end{tabular}
\end{table}

Given that the NCSBE does not release the details of the process by which they remove duplicates or how they choose the most representative record of an individual, this application motivates our methodology. We will return to this example in Section \ref{sec:application}.

\subsection{Related Work}
\label{sec:related}

The most common approach for ER is the Fellegi-Sunter model based upon pairwise comparisons of common data fields that are used to estimate conditional probabilities of matches and non-matches \citep{fellegi_theory_1969, sadinle_generalized_2013}. A drawback of these and other related probabilistic approaches \citep[e.g.,][]{larsen_2001, HofRavelliZwinderman17} is the lack of natural uncertainty quantification in the linkage structure, limiting natural error propagation to the downstream task. This particular limitation has led to many recent developments of Bayesian ER approaches \citep{tancredi_hierarchical_2011, zhao_bayesian_2012, gutman_2013, sadinle_detecting_2014, steorts_entity_2015, steorts_bayesian_2016, zanella_flexible_2016, sadinle_bayesian_2017, marchant2019d, mcveigh_scaling_2020}. 

While the canonicalization methods that we explore in this work are not tied to a particular ER model, we employ the approach of \cite{marchant2019d} to obtain posterior samples of the linkage structure (see Sections~\ref{sec:simulated} and \ref{sec:application}). In this model, ER is framed as a bipartite matching problem that links records to (unknown) latent entities. Although most Bayesian ER methods are flexible and can propagate error, they are known to suffer from scalability issues in realistically sized databases ($n \gtrsim 10^4$ records). Scalability has been addressed previously by splitting the data using a variable thought to be relatively clean \citep[\emph{blocking}, see ][]{christen_data_2012, steorts_comparison_2014} and applying a Bayesian ER model to each subset separately \citep{tancredi_hierarchical_2011, steorts_bayesian_2016, murray2016probabilistic, sadinle_detecting_2014, sadinle_bayesian_2017}. However, since the blocking and ER task are not modeled jointly, the error cannot be quantified exactly. These limitations are addressed in \cite{marchant2019d}, which is a scalable model that propagates uncertainty from blocking and ER simultaneously. 


The earliest proposals of canonicalization were deterministic, rule-based methods, which were application-specific and fast to implement \citep{cohen2005incremental}. Other existing literature involving probabilistic approaches commonly assumes the availability of training data in order to select the canonical data set. This assumption has lead to several optimization and semi-supervised methods for finding the most representative or canonical records \citep{yan1999conflict, bohannon2005cost, culotta2007canonicalization}. Compared to the existing literature, the canonicalization methods that we explore in this work do not rely on any training data (which can be expensive or difficult to obtain), making them fully unsupervised. Additionally, one of our proposed methods fully exploits the uncertainty quantification properties of the Bayesian framework to propagate the ER error to the downstream task (see Section~\ref{sec:canonicalization}). For a review of canonicalization and data fusion techniques, see \cite{bleiholder2009data}.

Recent work on the relationship between ER and downstream tasks commonly makes assumptions that remove the need for canonicalization. In fact, some downstream tasks have no need for a canonicalization stage because they use only functions of the linkage structure, e.g., population size estimation \citep{tancredi_hierarchical_2011, tancredi2020unified}. For those downstream tasks that do require the linked data, the literature can be classified into two main frameworks -- single and two-stage approaches. 

Single-stage approaches build one joint model for the ER and the downstream task, while two-stage approaches treat each model separately. 
Single-stage approaches for regression and classification have been limited mainly to linking two databases and do not easily generalize beyond this framework \citep{DalzellReiter18, steorts2018generalized, gutman_2013, HofRavelliZwinderman17}. Moreover, they require knowledge of the model specification up front such that if an additional downstream task is required (after the single-stage joint approach has been fitted), the linkage would need to be repeated in a new joint model for valid inference. Because ER is the most computationally costly piece of the joint model, this can be an argument against using a single-stage approach. 

Performing ER and the downstream task in a two-stage approach allows for specification of downstream models post-linkage, which can expand the computational feasibility of the methods at the cost of no longer having the downstream model inform the linkage. Most of the two-stage literature joins two databases and often assumes that the error from the ER task occurs only in the response variable \citep{lahiri_2005, kim12, goldstein12, hof12, chambers19}. In a more general setting where more than two databases are considered or where duplication and linkage errors can affect both response or explanatory variables, there is a dearth of literature. Our proposed work on canonicalization is intended to bridge the gap between ER and the downstream task in this more general setting.

{\bf Our Contributions:} The main contribution of this work is to provide a practical approach to downstream inference with linked data while maintaining principled error propagation under mild conditions. We propose a scalable, multi-stage approach, where the ER error is propagated to the downstream task via a Bayesian canonicalization stage. A central advantage of the proposed methodology for canonicalization is its generality -- it is applicable in the most general data scenario, one in which we have any number of databases to link and duplication can occur in all downstream variables. Our proposed canonicalization methods can be used following any ER method that produces a partition of the records, and prior to any downstream task. With this work, our goal is to make a step toward the formalization of a process that historically has been an ad hoc procedure. We aim to provide guidance for analysts and statisticians who are currently working with linked data on how they can achieve appropriate inference when working with less than perfect and noisy data sets.


The remainder of the paper is structured as follows. Section~\ref{sec:canonicalization} defines a \emph{canonical record} and \emph{canonical data set}, proposes five unsupervised methods of canonicalization, and describes the computational cost of our recommended method. 
Sections~\ref{sec:simulated} and \ref{sec:application} illustrate the proposed multi-stage approach on simulated data and the NCVD, respectively. Section~\ref{sec:discussion} provides a discussion, including advice for practitioners and directions for future work. We provide further details and a reproducible code base 
in the supplementary materials.

\section{Canonicalization} 
\label{sec:canonicalization}

We provide general definitions of a \emph{canonical record} and \emph{canonical data set} before detailing five methods for canonicalization.

\subsection{Notation}

Consider a collection of $T$ databases indexed by $i$, each with $n_i$ records (rows) indexed by $j$, and $p$ fields (columns) indexed by $\ell$. For instance, the NCVD data is a collection of $T = 5$ databases where each record is comprised of data fields such as first and last name, race, sex, age and political party (see Table \ref{tab:ex-records}). \cite{marchant2019d} assumes each record $(i,j)$ links to a single entity, denoted by $\lambda_{ij}$, from a fixed population of entities indexed by $e \in \{1, \dots, E\}$. For example, the five records in Table \ref{tab:ex-records} presumably link to the same individual voter. Denote the value of the $\ell$-th field for record $(i,j)$ by $x_{ij\ell}$ and the collection of all field values for all records to be linked is then denoted $\boldsymbol x = \{x_{ij\ell}: i = 1, \dots, T; j = 1, \dots, n_i; \ell = 1, \dots, p\}$. We assume that the fields $\ell$ can be split into those that are used for ER, $\boldsymbol x^a = \{x_{ij\ell}: i = 1, \dots, T; j = 1, \dots, n_i; \ell \in 1, \dots, p \text{ such that field } \ell \text{ is aligned in all databases}\}$, and those that are used for the downstream task, $\boldsymbol x^d$. Thus, $\boldsymbol x = \{\boldsymbol x^a, \boldsymbol x^d\}$. In the NCVD application (see Section \ref{sec:application}), $\boldsymbol x^a$ corresponds to the fields of first and last name, sex, and race, while $\boldsymbol x^d$ comprises age, ethnicity, and party affiliation. 

The observations of the aligned fields $x^a_{ij\ell}$ are assumed to be noisy observations of the true field value $z_{\lambda_{ij}\ell}$ of the linked entity. The linkage structure is defined as $\boldsymbol \Lambda = \{\lambda_{ij}: i = 1 \ldots T; j = 1 \ldots n_i\}$, where $\lambda_{ij}=e$ means that record $(i,j)$ is linked to the $e$ latent entity. If two records $(i, j)$ and $(i', j')$ refer to the same entity, then $\lambda_{ij} = \lambda_{i'j'}$ must hold. Under this framework, the linkage structure $\boldsymbol \Lambda$ induces a partition of the data into clusters where the number of clusters in the partition corresponds to the number of unique values of $e$ observed in the linkage structure. Let $C \equiv C(\boldsymbol \Lambda)$ be the set of clusters of records resulting from the linkage structure $\boldsymbol \Lambda$, assumed to fully partition the data $\boldsymbol x$. For simplicity of notation we will often drop the functional notation for $C$.

Note that when ER is performed with all the variables used for the downstream task, the latent values, $z_{\lambda_{ij}\ell}$, can be used as a representation of the truth in a downstream task. This requires all variables used in the downstream task (e.g., explanatory and response variables $\boldsymbol x^d$) to be also used as linkage variables (i.e., be included in $\boldsymbol x^a$), otherwise the latent values, $z_{\lambda_{ij}\ell}$, would not be available for those variables. However, it is a fundamental assumption in ER models that all fields in the linkage model are independent, which is not the case for the downstream variables. The effect of breaking this independence assumption has not been investigated in the literature. Furthermore, it may not be advisable to use downstream variables as linkage variables due to known noisiness or unreliability in the data (e.g age or party affiliation in Table \ref{tab:ex-records}). Additionally, since the use of canonicalization does not require all downstream variables to be known prior to the linkage, the ability to perform exploratory analyses is not limited \citep{tukey1977exploratory}. Thus, canonicalization remains a necessary step in many analysis pipelines. 

\subsection{Definitions}

The definitions of a canonical record and canonical data set are presented below.
\begin{definition}\label{def:can-record}
For each cluster $c \in C$, the canonical record $\boldsymbol r_c$, is defined as a function of the cluster and some known parameters $\boldsymbol \theta$, $r_c = \psi_{\boldsymbol \theta}(c, \boldsymbol x_c)$, where $\boldsymbol x_c = \{x_{ij\ell}: (i, j) \in c; \ell = 1, \dots, p\}$ denotes the record values in cluster $c$.
\end{definition}

\begin{definition}\label{def:can-data}
A canonical data set $\boldsymbol r$, is obtained by applying the canonicalization function $\psi_{\boldsymbol \theta}$ over each cluster that results from $\boldsymbol \Lambda$, that is $\boldsymbol r = \{\boldsymbol r_c: c \in C\}$.
\end{definition}

Definitions \ref{def:can-record} and \ref{def:can-data} include the record selection canonicalization methods proposed by \cite{culotta2007canonicalization} for the case where $\psi_{\boldsymbol \theta}$ is the minimum average edit distance between strings as a function of fixed or learned costs. However, our definition is more general as it includes categorical, ordinal and numerical fields, as well as unsupervised functions $\psi_{\boldsymbol \theta}$. 

\subsection{Point-Estimate Canonicalization}
\label{sec:point-estimate}

In this section, we assume that an ER task has been performed as to provide a point estimate $\hat{\boldsymbol \Lambda}$ of the linkage structure $\boldsymbol \Lambda$. In practice, the point estimate could result from any ER model that results in a partitioning of records. We propose three unsupervised methods for performing canonicalization based on $\hat{\boldsymbol \Lambda}$. 

First, we define \emph{random canonicalization}, which will serve as a baseline for comparison. For each cluster $c \in C$, choose the canonical record $\boldsymbol r_c$ randomly,
$$
\boldsymbol r_c = \{\boldsymbol x_{ck}: k \sim \text{Categorical}(\{(i,j) \in c\}, \boldsymbol \theta)\},
$$
where $\boldsymbol x_{ck}$ denotes the $k^{th}$ record in $\boldsymbol x_c$ and $\boldsymbol \theta$ represents the vector of selection probabilities for the records $(i,j)$ in cluster $c$.

\emph{Composite canonicalization} is defined as an aggregate record that includes information from each linked record in the cluster,
$$
\boldsymbol r_c = \{r_{c\ell}: \ell = 1, \dots, p\}, \quad r_{c\ell} = \overline{f}_{\ell}(\{x_{ij\ell}:(i,j) \in c\}),
$$
where $\overline{f}_{\ell}$ is a an aggregation function for each column, $\ell = 1, \dots, p$. The form of aggregation depends on the column type and can be weighted by some prior knowledge of the data sources when available. 
Composite canonicalization alters the original data values for records with duplicates (unless they are exact duplicates), which can heavily affect inference results for some downstream tasks. 


\emph{Minimax canonicalization} is a point-estimate canonicalization method designed to choose a canonical record that ``most closely captures" the underlying true unknown (latent) entity. We propose selecting the record whose farthest neighbor within the cluster is closest, where ``closeness" is measured by a pairwise record distance function, denoted by $d(\cdot, \cdot)$. 
We define the canonical record as the record $\boldsymbol r_c$ within each cluster $c$ such that
\begin{equation}\label{eq:minimax}
\boldsymbol r_c = \arg \min\limits_{(i, j) \in c}\max\limits_{(i', j') \in c} d(\boldsymbol x_{ij}, \boldsymbol x_{i'j'}), \quad c \in C
\end{equation}
The result is a set of representative records, one for each latent entity, that is central to the other records in each cluster. Many distance functions can be used depending on the context of the problem and have a critical role in determining the resulting canonical data set. One potential distance function is provided in Appendix A.1. Ties in maximum record distance within the cluster can be handled multiple ways, depending on the computational constraints and anticipated number of ties. One option is to select the record that has the closest farthest neighbor (minimax record distance) when compared to all other records in the data set (not within the cluster) that match on categorical variable levels. Another is to simply break the tie randomly. If tied records are identical, we select the record with the lowest index.  

We note that Bayesian models for ER are commonly sensitive to choice of hyperparameters, indicating a need for knowledge of the data collection process \citep{steorts_bayesian_2016, steorts_entity_2015, sadinle_bayesian_2017, aleshin2021multifile} and the choice of distance function in the minimax canonicalization method is similarly critical to its success.

\subsection{Posterior Canonicalization}
\label{sec:posterior}

We now propose two alternative approaches that utilize the marginal posterior distribution of the linkage structure, $P(\boldsymbol \Lambda | \boldsymbol x)$, to inform canonicalization using the probability that each record in the data is canonical, $P((i,j) \in \boldsymbol r | \boldsymbol x)$.
To estimate this probability, let $C^{(m)}$ represent the partition of the data into clusters for iteration $m \in \{1,\ldots,M\}$ of the Markov chain Monte Carlo (MCMC) samples from $P(\boldsymbol \Lambda | \boldsymbol x)$. We apply the following procedure,
\begin{enumerate}
\item For each $m$ and $c \in C^{(m)}$, find the canonical records, $\boldsymbol r_c^{(m)}$ based on minimax canonicalization defined in Eq.~(\ref{eq:minimax}).
\item Compute the \emph{posterior canonicalization (PC) weights} that approximate the posterior probability of each record $(i,j)$ being selected as a canonical record,
\begin{equation}\label{eq:PC}
\hat{p}_{ij} \equiv \frac{1}{M}\sum_{m=1}^{M} \mathbb{I}((i,j) \in \boldsymbol r^{(m)}) \approx P((i,j) \in \boldsymbol r | \boldsymbol x),
\end{equation}
where, $\boldsymbol r^{(m)} = \{\boldsymbol r_c^{(m)}: c \in C^{(m)}, m = 1,\ldots,M\}$.
\end{enumerate}
Note that the linkage uncertainty is captured by the (possibly) different partitions, $C^{(m)}$, obtained from the MCMC samples of $P(\boldsymbol \Lambda | \boldsymbol x)$. The PC weights described in Eq.~(\ref{eq:PC}) provide a vehicle through which to pass linkage uncertainty to the downstream task. 


\label{sec:PCweighted}

To incorporate the ER uncertainty in a downstream task via the PC weights $\hat{\boldsymbol p} = \{\hat{p}_{ij}: i = 1, \dots, T; j = 1, \dots, n_i\}$, we take advantage of the multitude of methods available for incorporating survey weights into downstream analyses \cite[e.g.,][]{little1991inference,pfeffermann1993role}. For illustration, assume the downstream task is a linear model of the form
\begin{align}
Y | \boldsymbol \beta, \Sigma_y, \boldsymbol X \sim \text{MVN}(\boldsymbol X^T\boldsymbol \beta, \Sigma_y), \label{eq:downstream} 
\end{align}
where the response variable is represented as $Y$ and explanatory variables $X$. The \emph{PC weighted} method incorporates the uncertainty from both entity resolution and canonicalization as the weights in weighted linear regression, where
\begin{equation*}
\Sigma_y = Q_y\sigma^2 \quad \text{and} \quad Q_y^{-1} = \text{diag}(w_{11}, \dots, w_{1n_{1}},\dots,w_{T1},\dots,w_{Tn_{T}}),
\end{equation*}
and $w_{ij} = \hat{p}_{ij}$ defined to be the PC weight for record $(i,j)$ from Eq.~(\ref{eq:PC}). Note this is not a formal canonicalization method as defined in Definition \ref{def:can-data} given that all the original records are passed to the downstream task along with their respective PC weights, but it does allow for proper accounting of uncertainty.

To construct a canonical data set based on the PC weights, we use thresholding. The \emph{PC threshold canonicalization} method can be implemented similarly for the downstream task described in Eq.~(\ref{eq:downstream}), where now
$$
w_{ij} = \boldsymbol I \{\hat{p}_{ij} \ge \tau_{PC}\}
$$
for each record $(i,j)$. A value of $\tau_{PC} = 0.5$ is a natural choice because it results in canonical records having posterior probability greater than $0.5$ of being representative. We describe one way to validate the choice of threshold $\tau_{PC}$ in Section~\ref{sec:application}. Note that the PC threshold method can result in multiple (or zero) records being selected as canonical for a cluster.

\subsection{Computational Complexity} 
\label{sec:complexity}

The posterior canonicalization approach allows us to deliver a method that is still feasible even for large data sets. To see this, let $|c|$ denote the size of each cluster $c \in C^{(m)}$, for $m = 1, \dots, M$ where $M$ is the number of posterior samples of the linkage structure available and $C^{(m)}$ is the resulting cluster from $\boldsymbol \Lambda^{(m)} \sim P(\boldsymbol \Lambda | \boldsymbol x)$. ER has been shown to exhibit the properties of \emph{microclustering}, where the cluster sizes grow sub-linearly as the size of the data increases \citep{zanella_flexible_2016, betancourt2020random}. Given the expected sub-linear growth of the cluster sizes, it is common to assume that  
\begin{align}
\label{eqn:sublinear}
\frac{1}{n}\max\limits_{\substack{c \in C^{(m)} \\ 1 \le m \le M}} |c| \rightarrow 0 \text{ as } n \rightarrow \infty,
\end{align}
for $n = \sum_{i=1}^T n_i$, the total number of records in all $T$ databases. Thus, the computational complexity of our proposed posterior canonicalization methodology (with randomly broken ties) is $\left(\max_{c, m} |c|\right)^2$, sub-quadratic with respect to $n$. For proof of this assertion, see Appendix A.2. We make no statement about the computational feasibility of other methods, rather we simply note that the computational complexity of the posterior canonicalization method is not prohibitive of larger analyses.

\section{Simulation Study}
\label{sec:simulated}

Given that the true underlying relationship between variables is not available in any real data application, we construct simulated data sets that contain this information as well as the true records. The aim of this simulation study is to provide an overview of the performance of different canonicalization methods and their effect in a downstream task. We frame the simulation study to perform the downstream task of linear regression. 
The simulated data sets are generated using the GeCO tool \citep{tran2013geco} and three different levels of noise in the relationship between predictors and response variables through Gaussian noise, where $\sigma = 1, 2, 5$. Each of the three data sets contain a total of 500 records, 30\% duplication, and the maximum number of duplicates of each record is $5$. Each data set contains the following fields: first name, last name, birth date, sex, education level, income (in 1000s), and blood pressure (bp). The bp variable was generated with a known (noisy) relationship to sex and income, and our goal is to assess how the fitted model is altered based on the canonical data set passed after the ER task. Additionally, we generated three sets of test records, of 500 records each, following the same data generation mechanism. 
We consider two data scenarios -- the most general case, in which all variables in the downstream task are subject to ER error and a more common case in the existing literature, where only the explanatory variables are subject to this error. For full details on the data generation process, refer to Section \ref{sec:appendix-data} in the supplemental material. 

%

First, we perform ER on the GeCO data sets using the Bayesian ER model of \cite{marchant2019d}. In this simulated data set, we have a true known unique identifier, and thus, we are able to ascertain the true performance of the ER task. To evaluate ER performance we compute the pairwise precision and recall \citep{christen_data_2012}, which correspond to the proportion of predicted links that are correctly estimated and the proportion of true links that are correctly estimated, respectively. In this case, the pairwise precision and recall for the point-wise linkage are 0.97 and 0.88, respectively\footnote{Note that while there are three GeCO data sets, they only differ in the noise level for the relationship between explanatory and response variables. Thus, the noise differences do not affect the record linkage, which is performed just once with the linkage variables.}. 
To obtain a point estimate for the linkage structure, we use the the shared most probable maximal matching set (SMPMMS) from \citet{steorts_entity_2015}. Alternative point estimates were considered, including decision theoretic approaches for optimal Bayesian estimation based on multiple loss functions (Binder's, the Normalised Information Distance, and the Variation of Information) for the space of partitions \citep{lau2007bayesian, wade2018bayesian, rastelli2018optimal, GreedyEPL}. 
The adjusted Rand indices when compared to the SMPMMS were all greater than $0.9722$, indicating robustness.



To assess the performance of canonicalization, we evaluate the distributional closeness of the canonical data set generated from canonicalization to the true records using an empirical Kullback-Leibler (KL) divergence metric \citep{wang2005divergence, silva2007universal}. Values closer to zero indicate closer distributions.
These results are based on simulating 100 data sets.
As expected, the empirical KL divergence values are slightly higher for the scenario where all the downstream variables are susceptible to linkage error ($[0.0036, 0.0085]$) compared to the error-free response setting ($[0.0036, 0.0071]$). For all levels of noise in the two error scenarios, the closest distributions to the truth are achieved through the minimax method ($[0.0036, 0.0045]$) with PC threshold method ($[0.0038, 0.0047]$) and PC weighted ($[0.0041, 0.0049]$) very close behind. We expect the composite method to perform poorly in terms of distributional closeness since this method will alter all records with a duplicate, and this is indeed the case ($[0.0057, 0.0085]$). 

Lastly, we examine the performance of the canonicalization methods through the downstream task of linear regression via three metrics: (a) bias in the fitted coefficients, (b) coverage of the credible intervals, and (c) mean square error (MSE) for test records from each of the models fitted with the canonical data sets. We fit the models with a Bayesian specification in \texttt{stan} \citep{rstanarm, rstan}, with Gaussian prior distributions for the parameters \citep{gelman2008, gelman_bayesian_2013} and parameters centered and scaled to be weakly informative. 
See Appendix B.3 for more detail. 

The linear regression model is specified as
\begin{align}
& Y_i | \boldsymbol \beta, \sigma, \boldsymbol X_i \stackrel{ind}{\sim} \text{N}(X_i^T\boldsymbol \beta, \sigma^2) \notag \\
&\beta_j | b_j \stackrel{ind}{\sim} \text{N}(0, b_j) \notag \\
&\sigma | a \sim \text{Exponential}(a). \label{eq:reg}
\end{align}
In this simulation, we are interested in assessing the effect of canonicalization method on inference in the relationship between blood pressure and sex and income. Due to the focus on effect of canonicalization method, rather than the effect of model misspecification in the downstream task, we have generated the data from a known (and correctly specified) model for the purposes of being able to evaluate bias and coverage after linkage for each of the proposed canonicalization methods. 

\begin{table}

\caption{\label{tab:m0-bias-coverage}\label{tab:m0-bias-coverage-supp} Mean and standard deviation (in parenthesis) for MSE, bias, and coverage of the 90\% credible interval for income for regression based on five canonicalization methods and the true data set for levels of noise $\sigma = 1, 2, 5$. Results are based on 100 simulated data sets.}
\centering
\begin{tabular}[t]{ccccccc}
\toprule
\multicolumn{1}{c}{ } & \multicolumn{3}{c}{Errors in All Downstream Variables} & \multicolumn{3}{c}{Errors in Explanatory Variables Only} \\
\cmidrule(l{3pt}r{3pt}){2-4} \cmidrule(l{3pt}r{3pt}){5-7}
Method & MSE & Bias & Coverage & MSE & Bias & Coverage\\
\midrule
\addlinespace[0.3em]
\multicolumn{7}{l}{\textbf{$\sigma = 1$}}\\
\hspace{1em}Random & 37.55 (10.67) & 0.0862 (0.06) & 0.31 & 38.72 (12.02) & 0.08262 (0.06) & 0.28\\
\hspace{1em}Composite & 24.95 (8.67) & 0.04961 (0.04) & 0.51 & 23.93 (8.19) & 0.05907 (0.05) & 0.39\\
\hspace{1em}Minimax & 7.28 (3.14) & 0.00083 (0.01) & 0.97 & \textbf{6.44 (2.65)} & \textbf{-0.00256 (0.01)} & 0.98\\
\hspace{1em}PC Weighted & \textbf{7.16 (2.17)} & 0.00971 (0.01) & \textbf{0.92} & 6.54 (2.00) & 0.00663 (0.01) & \textbf{0.97}\\
\hspace{1em}PC Threshold & 7.27 (3.08) & \textbf{0.00021 (0.01)} & 0.97 & 6.47 (2.60) & -0.00259 (0.01) & 0.98\\
\hspace{1em}True & 2.23 (0.08) & 0.00086 (0.01) & 0.92 & 2.24 (0.08) & -4e-05 (0.01) & 0.91\\
\addlinespace[0.3em]
\multicolumn{7}{l}{\textbf{$\sigma = 2$}}\\
\hspace{1em}Random & 45.51 (10.26) & 0.08976 (0.05) & 0.23 & 43.95 (11.10) & 0.08272 (0.05) & 0.31\\
\hspace{1em}Composite & 31.74 (8.98) & 0.0552 (0.05) & 0.48 & 32.12 (9.16) & 0.06139 (0.05) & 0.5\\
\hspace{1em}Minimax & 12.6 (3.04) & -0.00315 (0.02) & 0.92 & 13.25 (2.94) & 0.00121 (0.01) & 0.95\\
\hspace{1em}PC Weighted & \textbf{11.83 (2.09)} & 0.00746 (0.02) & \textbf{0.9} & \textbf{12.11 (2.14)} & 0.01101 (0.02) & 0.87\\
\hspace{1em}PC Threshold & 12.52 (3.05) & \textbf{-0.00269 (0.01)} & 0.95 & 13.1 (2.86) & \textbf{0.00076 (0.02)} & \textbf{0.92}\\
\hspace{1em}True & 8.36 (0.35) & 0.00048 (0.01) & 0.9 & 8.34 (0.31) & 0.00106 (0.01) & 0.92\\
\addlinespace[0.3em]
\multicolumn{7}{l}{\textbf{$\sigma = 5$}}\\
\hspace{1em}Random & 89.32 (11.75) & 0.09106 (0.06) & 0.4 & 89.81 (12.74) & 0.09 (0.07) & 0.38\\
\hspace{1em}Composite & 74.82 (9.57) & 0.0536 (0.06) & 0.67 & 74.12 (9.52) & 0.05039 (0.05) & 0.64\\
\hspace{1em}Minimax & 55.61 (3.80) & \textbf{-0.004 (0.03)} & 0.91 & 55.61 (3.62) & \textbf{-0.00336 (0.03)} & 0.95\\
\hspace{1em}PC Weighted & \textbf{47.64 (2.63)} & 0.00443 (0.03) & 0.87 & \textbf{47.61 (2.64)} & 0.0053 (0.03) & 0.86\\
\hspace{1em}PC Threshold & 55.67 (3.88) & -0.00468 (0.03) & \textbf{0.9} & 55.44 (3.64) & -0.00361 (0.03) & \textbf{0.9}\\
\hspace{1em}True & 51.09 (2.19) & -0.00233 (0.03) & 0.92 & 51 (1.85) & -0.00324 (0.03) & 0.93\\
\bottomrule
\end{tabular}
\end{table}

Table \ref{tab:m0-bias-coverage-supp} displays the MSE, bias, and coverage of the 90\% credible interval of the regression coefficient for the income variable. See Tables \ref{tab:m0-bias-coverage-intercept}--\ref{tab:m0-bias-coverage-sex-income} in Appendix B for results of the other model coefficients. The MSE, bias, and coverage results are promising for minimax, PC Threshold, and PC Weighted. However, PC Weighted is consistently the optimal choice (or close to it) in all three metrics, with the best predictive performance when compared to the other canonicalization methods.
The coverage and bias behavior of the PC weighted method is consistent for the other model terms, again see Tables \ref{tab:m0-bias-coverage-intercept}--\ref{tab:m0-bias-coverage-sex-income} in Appendix B for evidence. The inferential and predictive results highlight the advantage of error propagation from the ER phase of the PC weighted method for downstream tasks.

 
\section{Application to NCVD}
\label{sec:application}

We now return to the motivational example of the NCVD from Section \ref{voter-data}. Due to the size of the NCSBE snapshot data set (each snapshot contains between 8 million and 29 million voter registration records, resulting in more than 500 million total records), we limit our investigation to five snapshot data sets recorded on 2019-04-30, 2019-01-01, 2018-11-06, 2018-05-08, and 2018-01-01. Specifically, we consider Caswell County, which is located in North Central North Carolina and has nearly even membership among Democrats and Republicans, as well as a diverse population in terms of gender, age, and race \citep{ncsbe-caswell}. This results in 54,716 records to be linked. For further detail on how we curated the NCVD data set, see Appendix C.1.

Discrepancies with respect to the NCSBE voter identifiers have been previously noted, leading others to question whether the assignment of voter identifiers in each snapshot correspond to the same individual voter \citep{wortman2019record}. Due to the suspected issues with the state-assigned voter identities, it is impossible to assess the accuracy of our ER or canonicalization procedures. We first perform ER using the model of \cite{marchant2019d} and apply our proposed canonicalization methods before investigating how well we can predict party affiliation in a downstream classification task. 

We provide the pairwise precision and recall, 0.979 and 0.787 respectively, for the point-estimate $\hat{\boldsymbol \Lambda}$ of the linkage structure when compared to the NCSBE voter identifiers (which may or may not themselves be accurate). This indicates that our method of de-duplication is linking more entities than the NCSBE's procedure. Appendix C.3 provides trace-plots and hyperparameter values. The resulting 95\% credible interval for the number of unique voters in the data set is $[14,394, 14,590]$ and the point estimate is $\hat{n}=14,484$. According to the NCSBE, there are $14,740$ registered voters in Caswell County, which is in agreement with the comparative overlinking evidenced in the precision and recall. 

Next, we perform canonicalization using all methods in Section \ref{sec:canonicalization}. The top of Figure~\ref{fig:pc_weights1} displays the distribution of the resulting posterior canonicalization weights. Due to the large size of the data, we have broken ties randomly. Many records have weights around $0.2$, suggesting the presence of many ties, which may lead to the exclusion of voters from the canonical data set. In fact, the choice of $\tau_{PC} = 0.5$ results in only $4,919$ canonical records using the PC threshold method, which is far less than the point estimate of the number of unique individuals in the data ($14,484$).

\begin{knitrout}
\definecolor{shadecolor}{rgb}{0.969, 0.969, 0.969}\color{fgcolor}\begin{figure}
\includegraphics[width=\maxwidth]{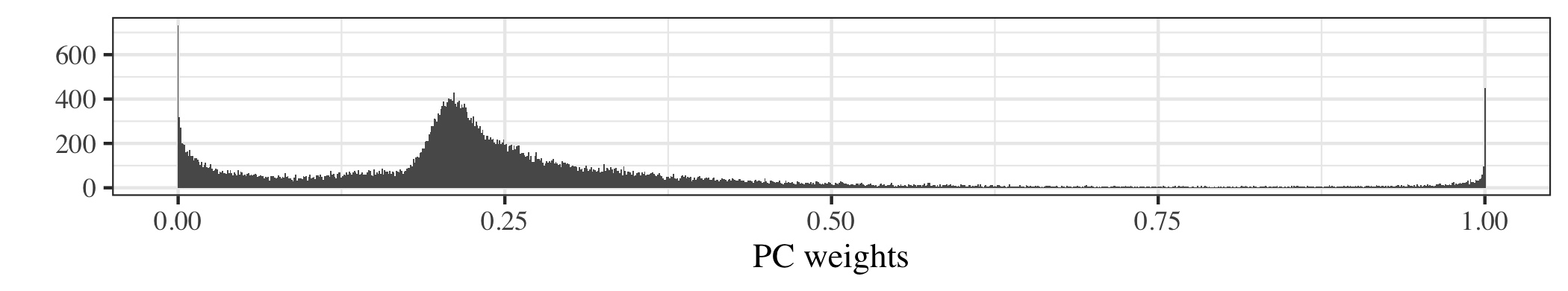} 
\includegraphics[width=\maxwidth]{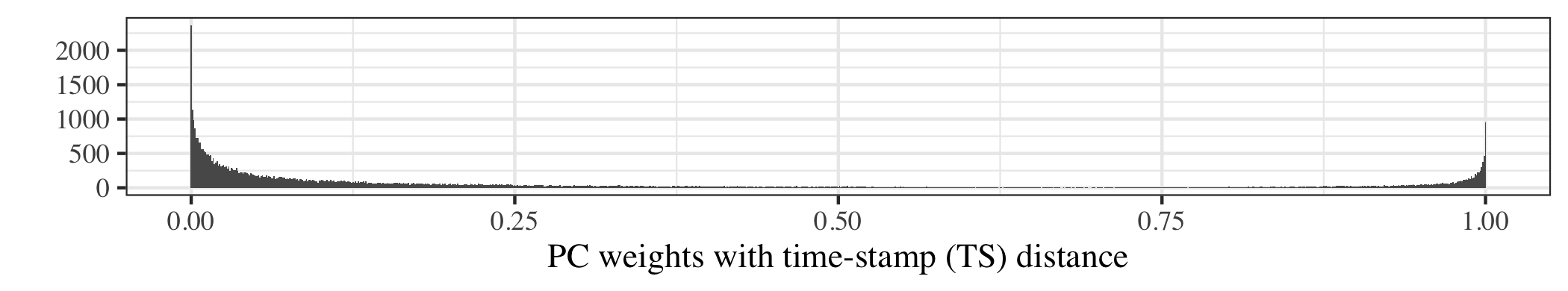} \caption{\label{fig:pc_weights1} (Top) Distribution of PC weights for Caswell NCVD. (Bottom) Distribution of PC weights using a distance function that includes time-stamps.}\label{fig:caswell-kl-compare-plot-1}
\end{figure}

\end{knitrout}

We overcome this issue by utilizing time-stamps in the distance function, which are available in this data set. 
This choice reflects our belief that the most recent record is likely the most accurate for this application. The bottom of Figure~\ref{fig:pc_weights1} displays the distribution of PC weights for a distance function that includes time-stamp information. Many records have PC weights close to $0$ or $1$, making the decision of excluding or including records in the canonical data set straightforward. Incorporating the time-stamp data in the distance function has eliminated the need for a robust tie-breaking procedure, allowing us to use the more computationally efficient version of posterior canonicalization.
The choice of $\tau_{PC} = 0.5$ results in $11,217$ canonical records using the PC threshold method, which is closer to the point estimate of the number of unique individuals in the data set ($14,484$).

We compute the empirical KL divergence between the canonical data sets of each of the proposed methods to the only notion of truth that we have -- the data set released for Caswell county by the NCSBE. KL values close to zero indicate distributional closeness between data sets. We reiterate that it is not known how the NCSBE performs ER or canonicalization, and thus, the quality of this data set is near impossible to ascertain. We find that the PC threshold TS method shows the lowest empirical KL divergence ($0.03$) when compared to the records released by the NCSBE. The PC weighted TS (Time-stamp) also provides a very low empirical KL divergence ($0.04$). On the other hand, the PC threshold method without temporal information provides the highest value ($0.1$). Certainly, the chosen value of $\tau_{PC}$ affects the result, but this behavior appears to be closely related to the large presence of record distance ties in this relatively clean data set. We emphasize this point, as it further highlights the importance of choosing a record distance function that has strong discriminatory power to distinguish between records and places the (believed) truth at the center of a cluster. From this outcome, we conclude that the PC methods that incorporate time-stamp information produce results that are most consistent with the de-duplication approach undertaken by the NCSBE.


Returning to the illustrative example of Table \ref{tab:ex-records}, we have appended two columns with the PC weights (both with and without time-stamp information) for each record in Table \ref{tab:ex-records} in Table \ref{tab:ex-records-weights}. PC weights above $\tau_{PC} = 0.5$ are bolded. Based on these weights, the canonical record that would be selected is the first entry in Table \ref{tab:ex-records-weights} using the PC threshold TS method. Based on the PC threshold (no time-stamp information), none of these records would be selected and this individual would be left out of future analyses. 

\begin{knitrout}
\definecolor{shadecolor}{rgb}{0.969, 0.969, 0.969}\color{fgcolor}\begin{table}

\caption{\label{tab:ex-records-weights}\label{tab:ex-records-weights} Five records that represent the same voter with their respective PC weights both with and without time-stamp information. PC weights above $\tau_{PC} = 0.5$ are bolded.}
\centering
\begin{tabular}[t]{llllllrr}
\toprule
Last & First & Race & Sex & Age & Party & PC Weight & PC Weight TS\\
\midrule
BAKER & MARK & WHITE & MALE & 42 & REP & 0.295 & \textbf{0.793}\\
BAKER & MARK & WHITE & MALE & 41 & DEM & 0.325 & 0.463\\
BAKER & MARK & WHITE & MALE & 41 & REP & 0.306 & 0.326\\
BAKER & MARK & WHITE & MALE & 40 & DEM & 0.124 & 0.05\\
BAKER & MARK & WHITE & MALE & 40 & REP & 0.232 & 0.053\\
\bottomrule
\end{tabular}
\end{table}

\end{knitrout}

We next consider the downstream task on Caswell County, where our goal is to be able to model the relationship between party affiliation and the following demographic variables (fields): sex, age, race, and ethnicity. We consider a logistic regression model fit using \texttt{stan} \citep{rstanarm} with Gaussian prior distributions for the parameters \citep{gelman2008, gelman_bayesian_2013} and parameters centered and scaled to be weakly informative (see Appendix C.2), 
\begin{align}
Y_i | \boldsymbol \beta, \boldsymbol X_i &\stackrel{ind}{\sim} \text{Bernoulli}\left(\frac{\exp\{\boldsymbol X_i^T \boldsymbol \beta\}}{1 + \exp\{\boldsymbol X_i^T \boldsymbol \beta\}}\right) \notag \\
\beta_i | b_i &\stackrel{ind}{\sim} \text{N}(0, b_i). \label{eq:log_reg}
\end{align}

After completing both ER and canonicalization, we assess the performance of the downstream task by obtaining out-of-sample predictions of party affiliation from a recent snapshot of the NCSBE (2019-05-14) that was not included in either the ER task or the canonicalization tasks (test data set). Figure \ref{fig:caswell-auc} shows two methods for assessing the affect of canonicalization on the downstream task. On the left, we assess the predictive performance of the model fit on each canonical data set using the posterior test AUC. Based on these distributions, PC weighted, PC weighted TS, and PC threshold TS show the best predictive results. On the right of Figure \ref{fig:caswell-auc} we see the posterior predicted $Pr(\text{DEM} | \text{Male})$, holding all other fields fixed at a typical value in the test data set \citep{king2000making}. 
There is a clear difference in this relationship for  models fit on point-wise canonical data sets and posterior canonical data sets. This result in conjunction with the improved inference evidenced in Section \ref{sec:simulated}, indicates improved inferential performance via the natural error propagation from the posterior canonicalization methods.



\begin{knitrout}
\definecolor{shadecolor}{rgb}{0.969, 0.969, 0.969}\color{fgcolor}\begin{figure}
\includegraphics[width=.5\textwidth]{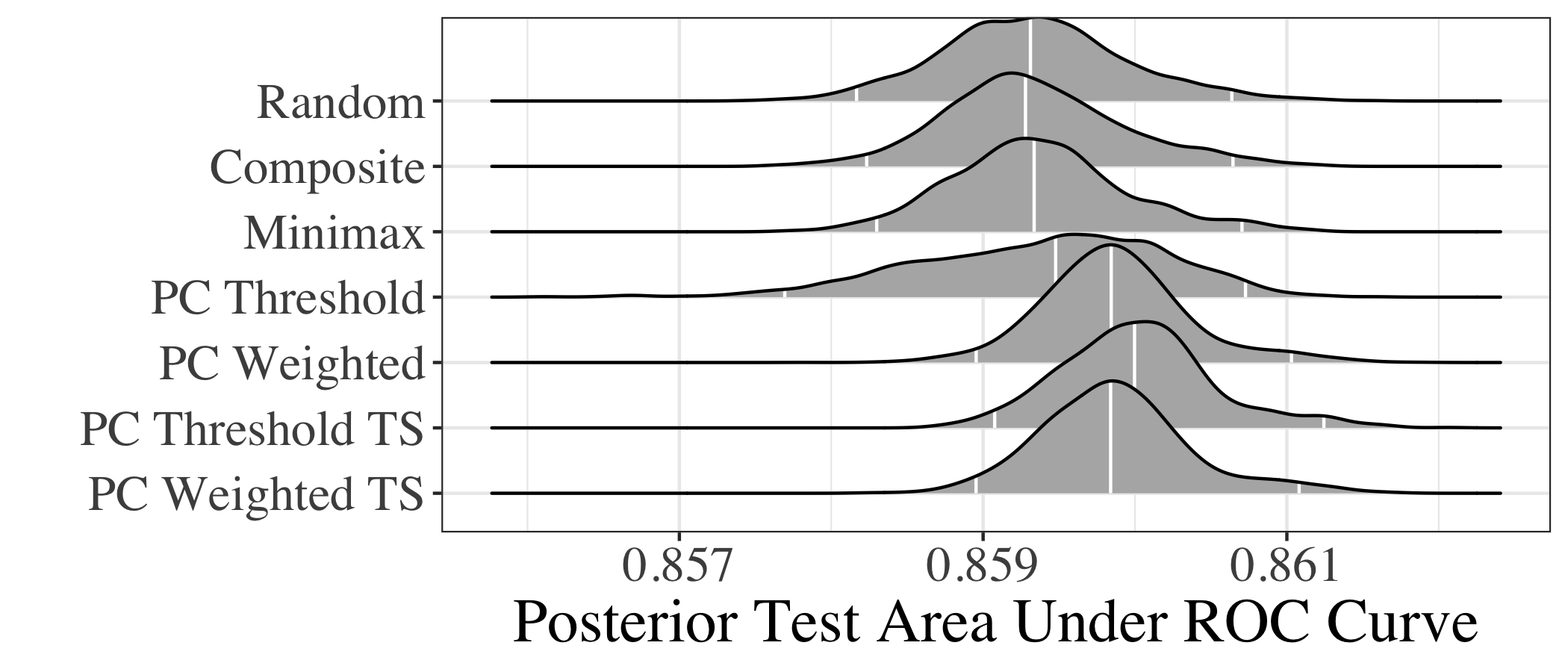} 
\includegraphics[width=.5\textwidth]{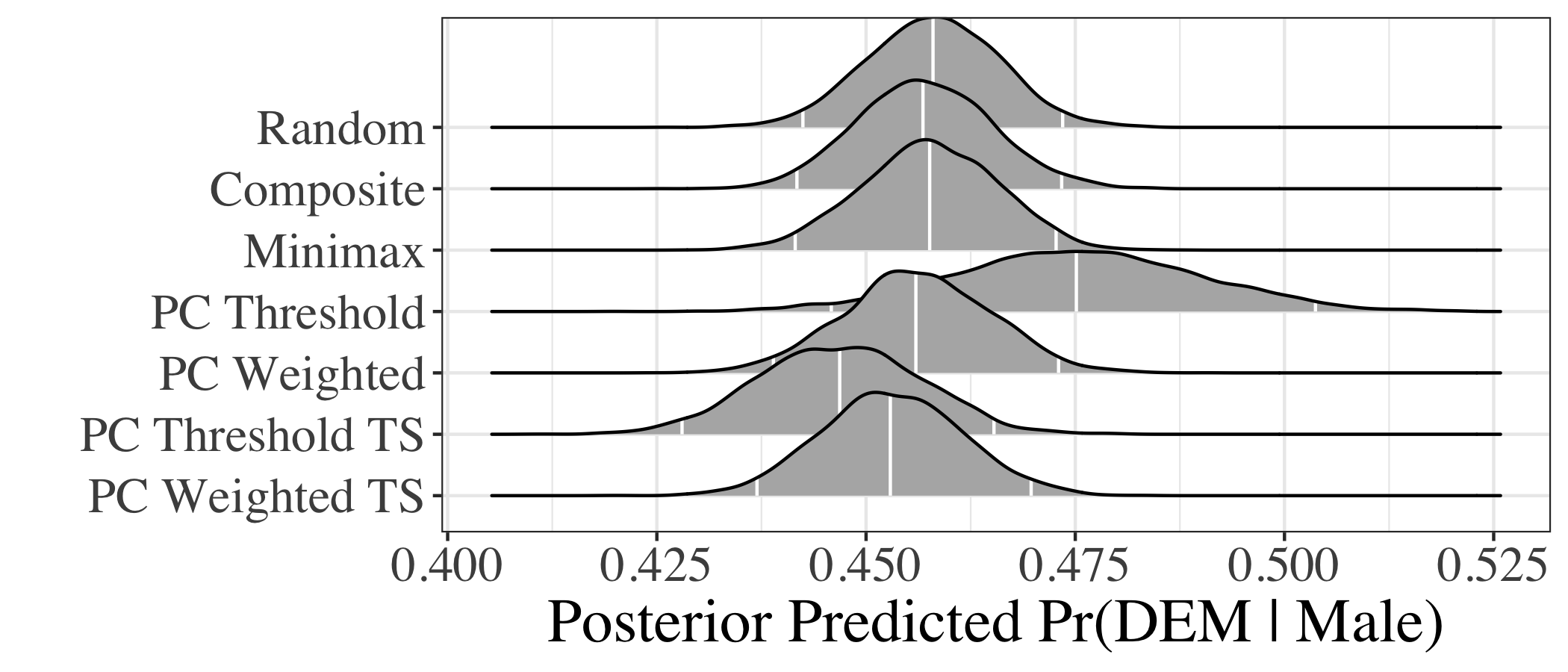} \caption{\label{fig:caswell-auc} (Left) Posterior Test Area Under ROC Curve for the logistic regression models fit after performing each type of canonicalization. PC weighted, PC weighted TS, and PC threshold TS show the best predictive results. (Right) Posterior predicted $Pr(\text{DEM} | \text{Male})$ holding all other features at a typical value in the test data set. There is a difference in the inferential relationship between outcome and predictors for models fit on point-wise canonical data sets versus posterior canonical data sets.}\label{fig:auc-plot-0}
\end{figure}

\end{knitrout}


\section{Discussion}
\label{sec:discussion}

In this paper, 
we have presented a practical approach to proper inference with linked data via canonicalization.
This approach allows error to propagate naturally into downstream analyses, such as prediction of voter affiliation in Caswell County or regression of blood pressure on sex and income. We have proposed several methods to find canonical records, including those based on point estimates and posterior distributions of linkage. Additionally, we have empirically shown the benefits of error propagation with posterior canonicalization through an inferential downstream task in simulated experiments, as well as a real data example. 


In general, we recommend the use of posterior canonicalization when two conditions are met -- Bayesian record linkage is used and a discriminatory distance function is available that places the true records central to their clusters. When Bayesian record linkage is not used, then we recommend minimax canonicalization with a robust tie-breaking procedure and when a discriminatory distance function is not available we recommend composite canonicalization. Finally, in the case where there is no reason to believe that the true records can be placed central within their clusters, then we are left with only random canonicalization. In this case, the use of a very informative distribution for records within cluster may be helpful. See Appendix D for a potential decision making process to determine which canonicalization method to use. 

The key advantages of our proposed methodology for canonicalization are generality and computational efficiency. The methods are applicable in general data scenarios with multiple databases where duplication can occur in all downstream variables at a relatively low computational cost. Canonicalization can be a crucial step that facilitates the transition between the ER stage and the subsequent downstream tasks in general applications with linked data. Future areas of research include investigating trade-offs in choice of ER model under simulated and real data and determining automated methods for choosing $\tau_{PC}$. It is also of interest to consider more downstream tasks, such as generalized linear models, small area estimation, and alternative classification methods.

\bigskip
\begin{center}
{\large\bf SUPPLEMENTARY MATERIAL}
\end{center}

The supplement contains additional details concerning data for the simulation study and NCVD application, additional results, diagnostics, and hyperparameters for the fitted models, as well as further advice for practitioners. The \texttt{R} code for reproducing the results is also available.

\bibliographystyle{apalike}
\bibliography{canonical}

\end{document}